\documentstyle[12pt,rotating]{article}
\begin{document}
\input epsf
\title{\begin{flushright}{\small HUB-EP-97/92}\end{flushright}\bigskip
\bf Van Alphen-de Haas effect  for dense cold quark matter 
in a homogeneous magnetic field}
\author{D. Ebert\footnote{e-mail:debert@qft2.physik.hu-berlin.de}\ \ 
and
A.S. Vshivtsev\footnote{On leave of absence from
Moscow Institute of Radioengineering, Electronics and Automatic
Systems, 117454 Moscow, Russia, e-mail: alexandr@vvas.msk.ru}\\
{\it Institut f\"ur Physik, Humboldt-Universit\"at zu Berlin,}\\
{\it Invalidenstrasse 110, D-10115, Berlin, Germany}}
\date{}
\maketitle
\begin{abstract}
We study the relativistic van 
Alphen--de Haas effect in cold dense quark 
matter in a homogeneous magnetic field which arises 
from oscillations in the effective potential of an 
underlying extended NJL-model.
Finally, we discuss the phase structure of the model in the
parameter plane $(\mu,H)$  and the solution of the mass gap equation.
\end{abstract}

\section{Introduction}
The study of dense matter 
phenomena is an important branch of physics. 
Most properties of matter are only weakly affected 
by changes in external parameters, such as temperature, 
chemical potential, electromagnetic field etc., when they vary within ranges
far from extreme values, i.e. under laboratory conditions. 
Therefore, each time when a small change in external 
parameters leads to a considerable response is 
remarkable and means a discovery of a significant effect. 
We can mention as examples the quantum 
Hall effect and oscillations of the magnetic moment 
predicted in 1930 by Landau \cite{lan} and experimentally 
observed by van Alphen -- de Haas  \cite{shen}.
Sixty years have passed since the discovery of this effect, 
but untill this day only quantum oscillations considered in 
the nonrelativistic case have been discussed extensively. 
Most attention of researchers dealing with oscillations 
of the magnetic moment is now focused on the relativistic 
problem since the results of these studies may be 
applied to cosmology, astrophysics and high-energy 
physics \cite{per,chak}.

The consideration of 
dense quark matter as described by a chemical potential leads to 
screening of quark charges and possibly to a 
restoration of chiral symmetry. In order to describe 
the corresponding chiral phase transition we must have a realistic model. 
It is well known that models of the Nambu-Jona-Lasinio 
(NJL) type with inclusion of background fields, 
temperature and chemical potential are a good 
laboratory for investigating the 
nonperturbative phenomenon of spontaneous breaking of 
chiral symmetry, as well as for describing the low-energy 
sector of quantum chromodynamics \cite{eb,vk,kl}

In this paper we shall in particular 
demonstrate the possibility of the existence of the van 
Alphen -- de Haas effect in quark matter with 
finite density in a homogeneous constant magnetic field. 
As model Lagrangian for light quarks we shall 
use the NJL model where quarks acquire a dynamical 
mass due to spontaneous breaking of 
chiral symmetry in the presence of a homogeneous magnetic background 
field. 

\section{Effective potential of the NJL model}

The Lagrangian of the NJL model with light quarks $u,d$ 
of the flavor group $SU(2)_f$ and the color group 
$SU(N_c)$ has the form (flavor and color indices of quarks will be suppressed)
\begin{eqnarray}
\label{eq.1}
L&=& \bar q i\gamma^\mu 
D_{\mu} q + \frac{G}{2N_c}[(\bar qq)^2+(\bar qi\gamma ^5\vec\tau q)^2],
\end{eqnarray}
where the covariant derivative is given by $D_\mu =
\partial_\mu -iQA_\mu,~~  
Q=diag(e_1,e_2)$, $G$ is a 
four-fermion coupling constant, and $\vec\tau$ 
are isospin Pauli matrices. As is well known, 
this model is symmetric under continuous chiral transformations. 

To investigate the properties of the vacuum 
of the NJL model it is convenient to employ 
instead of the quark Lagrangian (\ref{eq.1}) the quark-meson Lagrangian
\begin{eqnarray}
\label{eq.2}
L_\sigma&=&\bar q i \gamma^\mu D_\mu q - 
\bar q(\sigma + i \gamma^5 \vec\tau\vec\pi)q - \frac{N_c}{2G}\Sigma^2,
\end{eqnarray}
where $\Sigma^2=\sigma^2+\vec\pi^2$, 
which, due to the equations of motion for the 
fields $\sigma,\vec\pi$,  is equivalent to (\ref{eq.1}). 
It is convenient to consider the effective action of the model which is 
given by

$$
\exp\{iS_{eff}(\mu,\sigma,\vec\pi)\}=\int D 
\bar qDq\exp\left(i\int \left(L_\sigma+\mu\bar q\gamma_0q\right) d^4x\right),
$$

where
\begin{eqnarray}
\label{eq.3}
\frac{1}{N_c}S_{eff} = - \int d^4 x \frac{(\sigma^2 + \vec\pi^2)}{2G} - 
iTr\ln(i\gamma^\mu D_\mu +\mu\gamma_0- 
\sigma - i \gamma^5 \vec\tau\vec\pi),
\end{eqnarray}
and $\mu$ denotes the chemical potential.

Let us first recall the 
well-known properties of the model with vanishing chemical potential 
and vanishing background magnetic field. 
Assuming, as usual, that the fields $\sigma, 
\vec\pi$ do not depend on the space-time, and using the definition
$$
S_{eff} = - V_{eff}(\sigma,\pi)\int d^4 x,
$$
we obtain the effective potential
\begin{eqnarray}
\label{eq.4}
\frac{1}{N_c}V_{eff} = \frac{\Sigma^2}{2G} + 
4 i \int \frac{d^4p}{(2\pi)^4} \ln(\Sigma^2 - p^2) = \frac{V_0 (\Sigma)}{N_c}.
\end{eqnarray}
Switching in $V_{eff}(\sigma,\pi)$ to the Euclidean 
metric and introducing a  cut-off for the integration 
region $p^2 \le \Lambda^2$, we obtain
\begin{eqnarray}
\label{eq.5}
\frac{1}{N_c}V_{0}(\Sigma)&=& 
\Sigma^2 /2G -(8\pi^2)^{-2}\{\Lambda^4\ln(1+
\Sigma^2/\Lambda^2)+\cr &+&\Lambda^2\Sigma^2-
\Sigma^4\ln(1+\Lambda^2/\Sigma^2)\}.
\end{eqnarray}
The condition that the function (\ref{eq.5}) is stationary is
\begin{equation}
\label{eq.6}
\partial V_{0}(\Sigma)/\partial \Sigma = 0.
\end{equation}
Obviously, for $G < G_c = 4\pi^2/\Lambda^2$ 
equation (\ref{eq.6}) has no solutions 
except the trivial one $\Sigma=0$. On the other hand, 
for $G \ge G_c$ a nontrivial solution 
$\Sigma = \sigma_{0}$ does exist (we choose $\vec\pi_0 = 0$, 
which means that a pion condensate 
is absent), and the potential 
$V_0(\Sigma=\sigma_{0})$ has a global minimum 
at the point $\Sigma\ne 0$, signalling 
spontaneous breakdown of chiral symmetry and the appearance of 
a quark mass. 

In the following, we shall study a gas of quarks with 
non-zero chemical potential
$\mu$ in the presence of an external homogeneous 
magnetic field which is determined by the vector potential
$A_\mu = H\delta_{\mu2}x_1$. In this case the energy spectrum 
of quarks is given by
\begin{equation}
\label{eq.7}
E_{i,n}^2 =\Sigma^2 + p_3^2 + 2|e_i|Hn,
\end{equation}
where $n=0,1,2,...$  is the Landau quantum number. 
Let us next temporarily assume that, together with the chemical 
potential $\mu$, a heatbath with temperature $T$ 
acts on the system described by the NJL 
Lagrangian (\ref{eq.1}). In this case, to obtain the 
effective potential the following transformation of the 
integration variable $p_4$ must be made 
$p_4\to 2\pi(l+1/2)/\beta,~ 
l=0,\pm1,\pm2,...$, where $1/\beta = T$. 
After that the effective potential of our model can be written as
\begin{equation}
\label{eq.8}
\frac{1}{N_c}V_{\mu\beta}^{tot}(\Sigma,|e_i|H) = 
\frac{1}{N_c}V_{\mu\beta}(\Sigma) +\frac{1}{N_c}
V_{0\beta}(\Sigma,|e_i|H)+\frac{1}{N_c} \Delta V(\mu,\beta,\Sigma,|e_i|H),
\end{equation}
where $V_{\mu\beta}(\Sigma)/N_c$  
denotes the part of the effective potential with $\mu\ne 0$ and $H=0$,
and $V_{0\beta}(\Sigma,|e_i|H)/N_c$ is the part 
with $\mu = 0$, $H\ne 0$. Finally, $\Delta V(\mu,\beta,\Sigma,|e_i|H)$  
is responsible for the situation in 
which both $\mu$ and $H$ are non-zero. The third part 
of the effective potential is of particular interest here and takes the form
\begin{eqnarray}
\label{eq.9}
\frac{1}{N_c}\Delta V(\mu , \beta ,\Sigma ,|e_i|H) & = 
&- \sum _{i=1}^2 \frac{|e_i|HT}{\pi^2}\int _0^{\Lambda}\,
dp_3 \sum _{n=o}^{\infty} \alpha_n\{\ln [(1+
e^{-\beta(E_{i,n}+\mu)})\times\cr &&(1+e^{-\beta(E_{i,n}-\mu)})]\},
\end{eqnarray}
$$
\alpha_n=2-\delta_{n0}.
$$
Since we shall consider in this paper only the case of cold quark matter in
a magnetic background field $H$,
we let $T$ in (\ref{eq.9}) tend to zero which finally yields
\begin{equation}
\label{eq.10}
\frac{1}{N_c}\Delta V(\mu ,\Sigma ,|e_i|H)= 
- \sum _{i=1}^2 \frac{|e_i|H}{\pi^2}\int _0^{\Lambda}\,dp_3 
\sum _{n=o}^{\infty} \alpha_n \theta (\mu - E_{i,n})(\mu - E_{i,n})
\end{equation}
This representation will now be used for 
examining the phase structure of the NJL model.

\section{van Alphen - de Haas oscillations}

Next we will show that in the range of small magnetic fields 
the expression $\Delta V(\mu ,\Sigma ,|e_i|H)$ is an 
oscillating function of $|e_i|H$
which cannot be easily seen from the expression  
(\ref{eq.10}). Indeed, equation (\ref{eq.10}) is not 
quite convenient for extracting relevant physical informations. 
In order to resolve this problem, let us apply Poisson's summation formula
\begin{equation}
\label{eq.11}
\sum_{n=0}^{\infty}\alpha_n\Phi (n) = 2\sum_{n=0}^{\infty}
\alpha_n\int_0^{\infty}\Phi(x)\cos(2\pi kx)~dx
\end{equation}
to this equation. By integrating first over the momentum 
variables in (\ref{eq.10}), we find
\begin{eqnarray}
\label{eq.12}
\frac{1}{N_c}\Delta V(\mu ,\Sigma ,|e_i|H)&=
&-\frac{1}{\sqrt{2\pi}}\theta(\mu-\Sigma)\sum_{i=1}^{2}
\sum_{k=1}^{\infty}\left(\frac{|e_i|H}{\pi k}\right)^\frac{3}{2}
\{S_-(a_i,b_i)\cos \pi kb_i\cr \nonumber \\
&&  - C_-(a_i,b_i)\sin \pi kb_i\},
\end{eqnarray}
where
$$
S_-(a_i,b_i)=S(\pi ka_i)-S(\pi kb_i),~~~C_-(a_i,b_i)=C(\pi ka_i)-C(\pi kb_i).
$$
The functions $C(x)$ and $S(x)$ are given by Fresnel's integrals
\begin{eqnarray}
\label{eq.13}
{S(x)\choose C(x)}=\frac{1}{\sqrt{2\pi}}\int_0^x\frac{{\rm d}y}{\sqrt{y}}
{\sin y\choose\cos y},
\end{eqnarray}
and $a_i=\frac{\mu^2}{|e_i|H},b_i=\frac{(
\Sigma^2+\xi\Lambda^2)}{|e_i|H},\xi\in[0,1]$.  
Finally, after doing some calculations we obtain 
for the oscillating component of $\Delta V(\mu ,\Sigma ,|e_i|H)$ 
the expression
\begin{eqnarray}
\label{eq.14}
\frac{1}{N_c}\Delta V_{\rm osc}
(\mu ,\Sigma ,|e_i|H)&=&\frac{\mu \theta (\mu -\Sigma)}
{4\pi^\frac{3}{2}}\sum_{i=1}^{2}\sum_{k=1}^{\infty}
\left(\frac{|e_i|H}{\pi k}\right)^\frac{3}{2}
\{Q(\pi ka_i)\cos(2\pi k\omega_i+\frac{\pi}{4})
\nonumber \\
&& + P(\pi ka_i)\cos(2\pi k\omega_i-\frac{\pi}{4})\},
\end{eqnarray}
where $\omega_i=\frac{\mu^2-\Sigma^2}{2|e_i|H}$ are the 
oscillation frequencies. 

Note also that for $1 \ll x$ we have the representations
$$
P(x)=x^{-1}-3x^{-3}/4+... ,
$$
$$
Q(x)=-x^{-2}/2+15x^{-4}/8+... .
$$
Equation (\ref{eq.14}) is an exact expression 
for the oscillations of the effective potential 
of an ideal relativistic quark gas at $T=0$. 
The main difference of the relativistic case from 
the nonrelativistic one (discussed in QED by Landau in 
1930 (\cite{lan,shen,per,vsh}) is the oscillation 
frequency $\omega_i=\frac{\mu^2-\Sigma^2}{2|e_i|H}$ to 
be compared with the nonrelativistic expression 
$\omega_0=\frac{\Sigma(\mu-\Sigma)}{|e|H|}$. 
Further differences are the dependence of the dynamical quark mass
$\Sigma(\mu,H)$ on the chemical potential 
$\mu$ and the external background field $H$ (\cite{eb,kl,sh,vkm}). 
From  (\ref{eq.14})  we can easily derive the contribution 
to the magnetization
$$
\Delta M=-\frac{1}{\beta}\frac{\partial(\frac{1}
{N_c}\Delta V_{\rm osc}(\mu ,\Sigma ,|e_i|H))}{\partial H}
$$
which is obviously an oscillating function of 
$\omega_i$ (van Alphen - de Haas effect).

\section{Phase structure of the NJL model}

Next we shall examine the phase structure of the NJL model. 
For this purpose it is convenient to rewrite the third component 
of the effective potential given in (10) in a more useful for us form
which is completely equivalent to (\ref{eq.12}). To this end, let us 
integrate over the momentum variable which yields
\begin{eqnarray}
\label{eq.15}
&&\frac{1}{N_c}\Delta V(\mu,\Sigma,|e_i|H)=-
\frac{1}{\pi^2}\sum_{i=1}^2|e_i|H\sum_{n=0}^{\infty}
\alpha_n\theta(\mu-\sqrt{\Sigma^2+2|e_i|Hn})\cr
&&\times\left\{\mu\sqrt{\mu^2-\Sigma^2-2|e_i|Hn}+
(\Sigma^2+2|e_i|Hn)\ln\frac{\mu+
\sqrt{\mu^2-\Sigma^2-2|e_i|Hn}}{\sqrt{\Sigma^2+2|e_i|Hn}}\right\}.
\end{eqnarray}
It is convenient to consider in some detail 
the various terms in the gap equation
\begin{equation}
\label{eq.16}
\frac{1}{N_c}\frac{\partial V_{\mu}^{tot}(\Sigma,|e_i|H)}
{\partial \Sigma} = \frac{1}{N_c}\frac{\partial V_{\mu}
(\Sigma)}{\partial \Sigma} +\frac{1}{N_c}
\frac{\partial V_{0}(\Sigma,|e_i|H)}{\partial \Sigma}+
\frac{1}{N_c}\frac{\partial \Delta V(\mu,\Sigma,|e_i|H)}
{\partial \Sigma}=0
\end{equation}
which leads us to find some new properties 
concerning the phase structure of the model.
After some calculations we  find  for the first contribution
\begin{eqnarray}
\label{eq.17}
\frac{1}{N_c}\frac{\partial V_\mu(\Sigma)}
{\partial \Sigma}=\frac{\Sigma}{\pi^2}\left\{F(\Sigma)+ 
\theta(\mu-\Sigma)\left[
\mu \sqrt{\mu^2-\Sigma^2}-\Sigma^2 
\ln\frac{\mu+\sqrt{\mu^2-\Sigma^2}}{\Sigma}\right]\right\}
\end{eqnarray}
where $F(\Sigma)=\frac{\pi^2}{G}-\frac{\Lambda^2}{2}+
\frac{\Sigma^2}{2}\ln\left(1+\frac{\Lambda^2}{\Sigma^2}\right)$.

For the second term we have
\begin{eqnarray}
\label{eq.18}
\frac{1}{N_c}\frac{\partial {V_0(\Sigma,|e_i|H)}}{\partial 
\Sigma}=-\frac{\Sigma}{\pi^2}\left \{ 
\frac{1}{8\pi} \sum_{i=1}^{2}(2|e_i|H)
\left[\ln\frac{\Gamma\left({\frac{k_i}{2}}
\right)}{{\sqrt{2\pi}}}+ \frac{k_i}{2}-\frac{k_i-1}{2}\ln\frac{k_i}{2}\right]
\right\}.
\end{eqnarray}
where~ $k_i=\frac{\Sigma^2}{|e_i|H}$, and~ $\Gamma(x)$ is the 
gamma function.

Finally, the third contribution is given by
\begin{eqnarray}
\label{eq.19}
&&\frac{1}{N_c}\frac{\partial \Delta 
V(\mu,\Sigma,|e_i|H)}{\partial \Sigma}=
\frac{-2\Sigma}{\pi^2}\sum_{i=1}^2|e_i|H\sum_{n=0}^{\infty}
\alpha_n\theta\left(\mu-\sqrt{\Sigma^2+2|e_i|Hn}\right)\cr&&
\times\left\{\frac{-\mu}{\sqrt{\mu^2-\Sigma^2-2|e_i|Hn})}+\ln\frac{\mu
+\sqrt{\mu^2-\Sigma^2-2|e_i|Hn}}{\sqrt{\Sigma^2+2|e_i|Hn}}\right\}.
\end{eqnarray}
It follows from Eqs. (17)-(19)  
that the gap equation (16) takes the form
\begin{eqnarray}
\label{eq.20}
\Sigma\cdot(\Phi(\mu,\Sigma,|e_i|H))&=&
\Sigma\cdot(\phi_{\mu}(\Sigma) + 
\phi_0(\Sigma,|e_i|H) + \Delta \phi(\mu,\Sigma,|e_i|H)) = 0,
\end{eqnarray}
where the function 
$\Phi(\mu,\Sigma,|e_i|H) $ is represented as a sum of three terms 
corresponding to the R.H.S. of Eq. (16).

To investigate the phase structure 
of the model it is convenient to consider 
in the parameter plane $(\mu,h)$, $h=\sqrt{max(2|e_i|H)}$, 
the critical curves $\overline{\mu}_{(k)}(h)$ defined as solutions
of the equation  $\Phi(\mu,0,|e_i|H)=0 $ for the case 
where the dynamical quark mass $\Sigma$ vanishes and
chiral symmetry will be restored.
It is then useful to divide the part of 
the parameter plane in which $\mu\ge 0,h\ge 0$ into regions $\Omega_k$
\begin{equation}
\label{eq.21}
\Omega_k=\{(\mu,h):h\sqrt{k}\le \mu\le h\sqrt{k+1}\}
\end{equation}
limited from above by the critical curve $\bar\mu_{(k)}(h)$.
It is obvious that in the region $\Omega_0$ only the first 
term under the summation sign in the contribution 
from (\ref{eq.19}) is different from zero, in $\Omega_1$ 
the first and second terms are different from zero, and so on.
We can show that for fixed $h$ the dynamical quark 
mass $\Sigma(\mu)$ is monotonically decreasing in $\mu$. 
In particular, 
in discrete points $\mu_{k}=h\sqrt{k}$, 
k=1,2,..., its derivative turns out to have 
jumps $C_{k}(h)-C_{k-1}(h)~\ne~0$ where $C_{k}(h)$ is defined by
\begin{equation}
\label{eq.23}
C_k(h)=-\left(\frac{\frac{\partial\Delta\phi}{\partial\mu}}
{\frac{\partial\Delta\phi}{\partial\Sigma}}\right)_{\mu=\mu_k}, 
\end{equation}
and for all $k$ one has $C_k>C_{k+1}$ ($k=0,1,2,...$) and $C_k~<~1$.
Thus in the region
$0<\mu<\overline{\mu}_{(k)}(h)$ we obtain a nontrivial 
solution $\Sigma(\mu,|e_i|H)\ne 0$, whereas for 
$\mu>\overline{\mu}_{(k)}(h)$ the gap equation has only 
the trivial solution. The corresponding behaviour is shown schematically 
in the Fig.1.

\section{Summary}

In this work we have investigated dense 
cold quark matter in a homogeneous magnetic field 
using an extended NJL-model. Here, we were interested
in the case that chiral symmetry is broken so that quarks 
get a dynamical mass $\Sigma(\mu,|e_i|H)$. 
In particular, we have shown that the effective potential 
of the model is an oscillating function with 
frequency $\omega_i=\frac{\mu^2-\Sigma^2}{2|e_i|H}$.
In the case $\Sigma(\mu,|e_i|H)\sim\mu$ the relativistic 
frequency reduces to the well known nonrelativistic 
expression  $\omega_0=\frac{\Sigma(\mu-\Sigma)}{eH}$ 
\cite{lan,shen}. Finally, the phase structure of the NJL model 
in the parameter plane $(\mu,h)$
and the influence of the oscillations on the $\mu$-dependence 
of the dynamical quark mass
$\Sigma(\mu,|e_i|H)$ have been shortly discussed.

\section*{Acknowledgements}
One of the authors (A.S.Vsh.) gratefully acknowledges the support by DAAD and
the hospitality of the colleagues of the Particle Theory Group at the
Humboldt University extended to him during his stay there.

\vskip 5cm
\begin{figure}[hbt]

\centerline{\begin{turn}{-90}\epsfxsize=8cm
\epsfbox{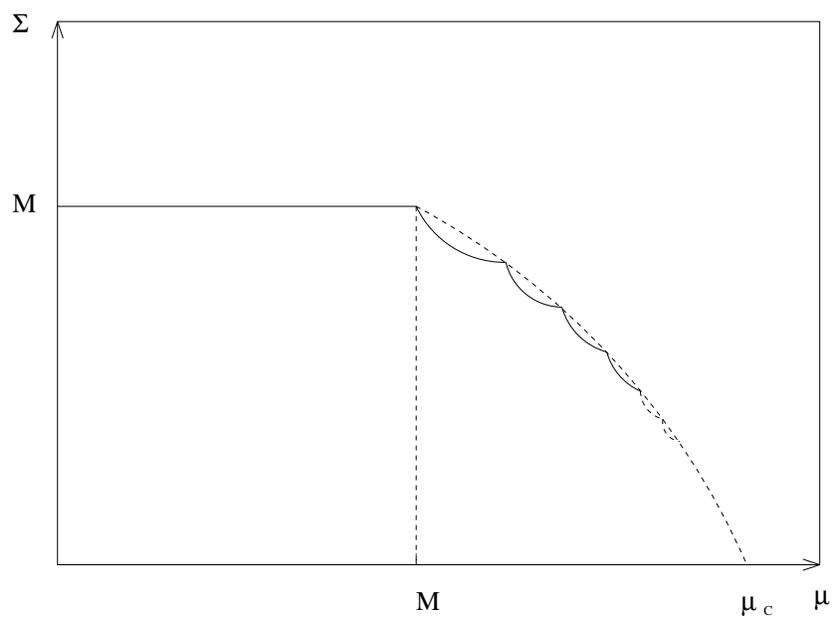}\end{turn}}

\caption{Schematical representation of the solution $\Sigma(\mu,|e_i|H)$ 
of the gap equation (\ref{eq.16}) as a function of $\mu$ for fixed $H$ 
with a given critical value $\mu_c=\bar\mu_{(k)}(h)$.}
\label{d}
\end{figure}

\end{document}